# Degeneracy of the 1/8 Plateau and Antiferromagnetic Phases in the Shastry-Sutherland Magnet TmB$_4$

Jennifer Trinh[1], Sreemanta Mitra[2], Christos Panagopoulos[2], Tai Kong[3], Paul C. Canfield[3], Arthur P. Ramirez[1]

[1]UC Santa Cruz, Santa Cruz, California 95064 USA
[2]Division of Physics and Applied Physics, School of Physical and Mathematical Sciences, Nanyang Technological University, 637371, Singapore
[3]Ames Laboratory, Iowa State University, Ames, Iowa 50011, USA

*Abstract*:

The 1/8 fractional plateau phase (1/8-FPP) in Shastry-Sutherland Lattice (SSL) spin systems has been viewed an exemplar of emergence on an Archimedean lattice. Here we explore this phase in the Ising magnet TmB$_4$ using high-resolution specific heat (*C*) and magnetization (*M*) in the field-temperature plane. We show that the 1/8-FPP is smoothly connected to the antiferromagnetic (AF) phase on ramping the field from *H* = 0. Thus, the 1/8-FPP is not a distinct ground state of TmB$_4$. The implication of these results for Heisenberg spins on the SSL is discussed.

*Keywords*: TmB$_4$, Shastry-Sutherland lattice, geometrical frustration, quantum magnetism, hysteresis, specific heat, magnetization plateau, SrCu$_2$(BO$_3$)$_2$

Magnetic systems with geometrically frustrated interactions[1,2] have produced a number of unconventional collective states, including spin ice[3,4] quantum spin liquid-like states[5], and of interest here, fractional magnetization plateaus[6]. Although plateaus often occur for short-range interactions on triangle-based structures, further-neighbor interactions can stabilize them on non-frustrating lattices, a compelling example of which is the Shastry-Sutherland lattice (SSL). The SSL, known for its exact ground state solution[7] is realized in $SrCu_2(BO_3)_2$ [8,9] as well as in $RB_4$, where $R$ is a rare-earth element[10-12]. These SSL-containing compounds exhibit plateaus in the magnetization ($M$) at rational fractions (e.g. 1/2, 1/3, 1/8) of the saturation value ($M_{sat}$).

While the experimental existence of plateaus in SSLs is firmly established, different theoretical descriptions have been proposed. The FPPs in $SrCu_2(BO_3)_2$ have been described as crystals of triplon (S=1) excitations[13-16] and crystals of S=2 triplon bound states[17,18]. Another approach maps spin operators to hard-core bosons and then to spinless fermions coupled to a Chern-Simons gauge field[19]. Recent work shows that, with small interaction anisotropy, the triplons themselves form topological bands with Chern numbers $\pm 2$ at low field (zeroth plateau)[20,21]. If true that topological, perhaps non-Abelian, excitations can exist in an FPP "vacuum", then the stability of the FPP itself should be established in order to distinguish between such excitations and fluctuations of phase. Among the FPPs that such theories need to replicate, the 1/8-FPP has presented a puzzle. While both the Heisenberg $SrCu_2(BO_3)_2$ as well as Ising $TmB_4$ exhibit this phase in both thermodynamic[8,9,22-30] and local[27,31] probes, it is not present in every theory. For example, some theories produce the 1/8-FPP for $SrCu_2(BO_3)_2$ [19,32-35], but the Chern-Simons mapping does not. One might argue that the observation of FPPs of similar fractions in both Heisenberg and Ising systems with different energy scales and ranges of interaction suggests a shared origin. Thus, in order to discuss the FPPs in a materials-agnostic way, it is imperative that their experimental stability be firmly established.

Here, we investigate the 1/8-FPP region in $TmB_4$ using both magnetization and specific heat ($C$). This metallic, tetragonal ($P4/mbm$), quasi-2D compound has Ising-like effective spins ($Tm^{3+}$, $J = 6$ non-Kramers doublet) interacting primarily via RKKY indirect exchange. The magnetic field ($H$) versus temperature ($T$) phase diagram has been studied with $M$[11,25,26,29], $C$[26,28], neutron diffraction[10,27], and charge transport[30], and the major features are shown in the inset of Fig. 1. The 1/8-FPP appears as a narrow region in the $H$-$T$ plane between $H = 1.40T$ and $1.75T$ and hysteresis in the value of the magnetization in the plateau region, as opposed to the location of its boundaries, is seen. In this region, Siemensmeyer et al.[27] and Wierschem et al.[29] observe $M/M_{sat}$ fractions of 1/7, 1/9, and 1/11, in addition to 1/8. By performing complementary measurements of $C$ and $M$ using the same $H$-$T$ sweep protocols, we address the thermodynamic nature of the 1/8-FPP. We find hysteresis in $C(H)$ around the FPP region, suggesting a dynamical origin of this phase. More importantly we find that, on approaching the 1/8-FPP region from $H = 0$, it is possible to enter this region from the AF state without crossing a phase line. This result suggests that the 1/8-FPP is either not symmetry-distinct from the AF phase or that the transition proceeds through other nearly-degenerate long-wavelength states. Such near-degeneracy of states may help to explain why the 1/8-FPP has been difficult to reproduce theoretically.

The 0.28 mg crystal used here was grown from solution using a technique described elsewhere[29]. Measurements of $M$ were obtained with a commercial SQUID magnetometer. For measurements of $C$, the sample was mounted with silicone grease on a small copper block, which became part of the addendum, and all such data were obtained using the thermal relaxation technique. Measurements of both $M$ and $C$ were performed with $H$ along the $c$-axis, i.e. normal to the 2D planes. In this direction, the demagnetization factor is $4\pi(0.15 \pm 0.02)$ for our sample whose $a$:$b$:c dimensions are 0.30:0.30:0.65 mm [36]. The protocol used for the measurements used to compare $M$ and $C$ was: 1) cool the sample to a temperature $T$ with $H = 0$; 2) ramp $H$ to 5T, where $M = M_{sat}$; 3) ramp $H$ to 2T; 4) obtain data (either $M$ or $C$) on ramping $H$ down; 5) ramp $H$ to 0 and back up to 1.4T; 6) obtain data on ramping $H$ up to 2T. By performing both $M$ and $C$ measurements on the same sample with the same protocol, the features associated with the FPP region can be faithfully compared.

In Fig. 1 we show $C(H)$ for different values of $T$ encompassing the entire phase diagram, shown in the inset. Additional points on the diagram came from measurements at fixed $H$, the data of which are available in the supplemental material. The ordering features are more clearly defined than in previous $C(H)$ measurements[28], suggesting high sample quality and intrinsic origins for the effects discussed here. In the following, we will focus on the region of the phase diagram enclosed by a dotted rectangle in the phase diagram.

In Fig. 2 we show $C(H)$ and $M(H)$ for six different temperatures encompassing the FPP. Similar to previous studies, we observe jumps in $M(H)$ centered at 1.40T and 1.75T. These jumps are spread out over small regions of *external* field $H$, and thus are consistent with first-order transitions as a function of the *internal* field, $H_i$. In such a case $\chi^{-1} = (\partial M/\partial H)^{-1} = N$, where $N$ is the demagnetization factor, in the regions where $M$ is increasing. Using $M_{sat} = 7\mu_B$, we find that $\chi^{-1} = 4\pi(0.166)$, which is within experimental error for the estimated demagnetization factor of our sample. This suggests that the narrow regions where $M$ is rapidly changing are mixed phases of the 1/8-FPP with the AF phase (0.138T < $H$ < 0.143T) and the ferrimagnetic (FI) phase (1.73T < $H$ < 1.81T). In the following, we present $M$ and $C$ as a function of external field, $H$. The demagnetization correction for $C$ itself will be greatest in the transition regions[37]. Such a correction is only meaningful, however, for a uniform $H_i$ and absent domain structure. As we show below, such assumptions are likely not valid in the 1/8-FPP region, and thus we will discuss $C$ as a function of $H$ with no loss of generality but realizing the observed peaks will likely be sharper when expressed in terms of $H_i$.

In Fig. 2, at the lowest and highest temperatures, we see two limiting behaviors. At 2.0K (Fig. 2f), the transition from AF to FI states proceeds via two distinct and nearly reversible transitions. The step in magnetization, from $M/M_{sat} \approx 0$ to $M/M_{sat} \approx 1/8$ in the 1/8-FPP at $H$ = 1.40T is accompanied by a corresponding peak in $C/T$, as expected for a thermodynamic transition. As alluded to above, a distinction should be drawn between a magnetization-reversal process, occurring for example in a hard ferromagnet, and a thermodynamic transition, i.e. one involving a thermodynamic number of spins. The former involves virtually no change in the local spin configuration and would not necessarily be accompanied by a corresponding peak in $C(H)$, whereas the latter involves re-configuration of local spin textures at inter-atomic spacings and would be signaled by a peak in $C(H)$. At 2.0K, $C(H)$ provides evidence that the $M(H)$ step is indeed a thermodynamic transition. Similarly, at $H$ = 1.75T, $M/M_{sat}$ jumps up to 0.5 on entering the FI state, accompanied by another peak in $C/T$. The signatures of these two transitions are virtually the same on increasing and decreasing magnetic field, indicating that microscopic statistical processes are governing the macroscopic response.

The behavior at $T$ = 4.5K (Fig. 2a) is qualitatively different from that at 2.0K. Whereas for increasing $H$, the ordering features in $C/T$ and $M/M_{sat}$ are observed only at $H \approx 1.75$T, for decreasing $H$, an additional jump down in $M/M_{sat}$ at $H$ = 1.47T is observed but *not* accompanied by a corresponding peak in $C/T$. Thus, while a plateau in $M(H)$ has developed, this behavior is not mirrored in *C*(*H*). We note that the temperature difference between 2.0K and 4.5K is large in relative terms and that for *H* = 1.3T (just below the 1.40T step) $M/M_{sat}$ = 8.2 × 10$^{-3}$ at 2.0K and 2.8 × 10$^{-3}$ at 4.5K. Thus, at 4.5K the number of spin-flip processes available for rearranging magnetic order is almost three times larger than at 2.0K.

We gain insight into the processes governing the transition out of the 1/8-FPP from the behavior between 2.5K and 4.0K, shown in Fig. 2 b-e. For field up-sweeps, the plateau value of $M/M_{sat}$ decreases from 0.114 (2.0K) to 0.027 (3.5K), a factor of 4.2. On field down-sweeps, however, the behavior is qualitatively different. In the same range of *T*, $M/M_{sat}$ changes from 0.134 (2.0K) to 0.154 (4.5K), only a 14% increase. Thus, the hysteresis loop in $M/M_{sat}$ opens up as temperature is *increased*, an effect that results primarily from the reduction of magnetization on field up-sweep. The data in Fig 2 d-f show that this reduction in $M/M_{sat}$ is accompanied by the vanishing of critical behavior in *C/T* on up-sweep at *H* = 1.40T. Whereas on down-sweep, $M/M_{sat}$ exhibits little change in the 1.40T jump over the entire temperature region, the critical response of *C/T* on down-sweep vanishes with increasing *T*. Thus, we see that different protocols by which the 1/8-FPP is approached yield qualitatively different pictures of its lower field boundary.

As seen in Fig. 2, the *C*(*H*) data change dramatically between 2.0K and 4.5K, suggesting the presence of a *T*-constant phase boundary. In order to define this phase boundary, we performed *C*(*T*) measurements at several values of fixed *H*, for both increasing and decreasing *T*, as shown in Fig. 3a and 3b, respectively. Indeed, we see sharp ordering features at T ~ 4.2K for *H*-values that bracket the 1/8-FPP. Both above (2.0T) and below (1.3T) the 1/8-FPP region, the peaks broaden into a short-range-order feature. Thus, we observe a *T*-constant phase boundary, not previously reported, that, along with the *H*-constant boundaries at 1.41T and 1.75T, fully delineate the 1/8-FPP region on *H* down-sweeps.

The phase boundaries obtained around the 1/8-FPP region as defined by *C*(*T*,*H*) are shown in Fig. 4. While the 1/8-FPP region is now well-defined, it is only bounded and distinct from the AF phase with decreasing *H*. When data are obtained on increasing *H*, a gap in the boundary is seen between *T* = 2.5K and *T* = 4.0K, allowing paths from the AF to 1/8-FPP regions without a thermodynamic ordering feature. Thus, it is likely that the presence other FPPs, indicated by the seemingly continuous reduction of *M* in this region on increasing *H* leads to the traversal of a sequence of nearly degenerate phases en route from the AF to the 1/8-FPP region. This situation is akin to the critical point of water, around which exist paths in the pressure-temperature plane that allow the conversion from gas to liquid without traversing a critical line. On sweeping down in field, the system must transform from the FI state in which ½ of the spins are fully aligned, to the 1/8-FPP. This dramatic reorientation of spins presumably creates the dynamic phase space for selecting the lowest energy 1/8-FPP.

A simple state-energy analysis reveals why the AF, 1/8-FPP, and other fractional FPPs are nearly degenerate between 1.41T and 1.75T. Following Tinkham's[38] treatment of the metamagnet $CoCl_2 \cdot 2H_2O$, and using the ordered patterns reported by Siemensmeyer et al.[27] for the AF and FI phases of $TmB_4$, we equate the $T = 0$ energy of the AF and FI phases at $H_{c1} = 1.5$T, and the FI and paramagnetic (fully polarized at $T = 0$) phases at $H_{c2} = 3.7$T. This yields values for the SSL nearest neighbor exchange interactions of $J_1 = 0.45$K and $J_2 = 1.25$K, assuming an effective spin of 6, producing an energy difference of $3.5k_B$ between the AF and FI states at $H = 0$, as shown in Fig. 3c. We know that the FI and AF state energies ($E$) must obey $dE/dH > 0$, and we make the reasonable assumptions *i*) that $E$(1/8-FPP) is greater than $E$(AF) at $T = 0$, $H = 0$; *ii*) that $E$(1/8-FPP) = $E$(AF) at $H = 1.41$T and; *iii*) that $E$(1/8-FPP) > $E$(FI) for $H > 1.75$T. These constraints dictate a $T = 0$ energy level diagram similar to that shown in Fig. 3c. Increasing $T$ will reduce the energy differences but not substantively change the constraint conditions, as evidenced by the negligible dependence of $H_{c1}$ and $H_{c2}$ on $T$. We see, therefore, that the difference in energy between the AF and 1/8-FPP is only a few tenths of a Kelvin, and cannot change significantly for different $dE/dH$ values, given the above constraints. The 1/8-FPP is created from the AF state by flipping 1/16 of the spins. The plateaus with $M/M_s < 1/8$ observed on up field sweeps are created with even fewer spin flips. For example, the plateau seen for $T = 3.0$ K in Fig. 2d has $M/M_s = 1/30$, which is obtained by flipping 1/60 of the spins. The near degeneracy of these states is shown schematically in the inset of Fig. 3c.

We have shown that the 1/8-FPP can be accessed via the AF state in a manner similar to the triple point of water. We have also shown that $M(T, H)$ in the 1/8-FPP region can adopt a seemingly continuous set of values in $H$ up-sweeps. These observations lead us to conclude that the 1/8-FPP is not a thermodynamically stable state, but rather a metastable variant on the AF state created on approaching the phase boundary to the 1/2-FPP. This result has important ramifications for the study of FPPs. First, it shows that an observed FPP should not be construed as a stable ground state of the system. Thus, the inability of the Chern-Simons mapping for $SrCu_2(BO_3)_2$ to reproduce the 1/8-FPP might not necessarily be a failure of the theory. Second, plateau phases need to be reconciled with the complete phase diagram. The type of study presented here will be difficult to perform on $SrCu_2(BO_3)_2$ given that the 1/8-FPP occurs at the high field of $H = 27T$, but the present study provides motivation to pursue such work. Theoretically, FPP systems are good candidates in which to conduct searches for topological, and possibly non-Abelian, excitations. The present work shows that, in order to even begin such a search, phase stability needs to be established.

This work was supported by the U.S. National Science Foundation grant number NSF-DMR 1534741 (J.T.), U.S. Department of Energy grant DE-SC0017862 (A.P.R.), Ministry of Education, Singapore MOE2014-T2-2-112 (S.M. and C.P.), the Singapore National Research Foundation, Investigatorship NRF-NRFI2015-04 (S.M. and C.P.). Work at Ames Lab (P.C.C. and T.K.) was supported by the U.S. Department of Energy, Office of Basic Energy Science, Division of Materials Sciences and Engineering. Ames Laboratory is operated for the U.S. Department of Energy by Iowa State University under Contract No. DE-AC02-07CH11358. We would also like to thank Sriram Shastry and Steve Simon for helpful insights.

Figures

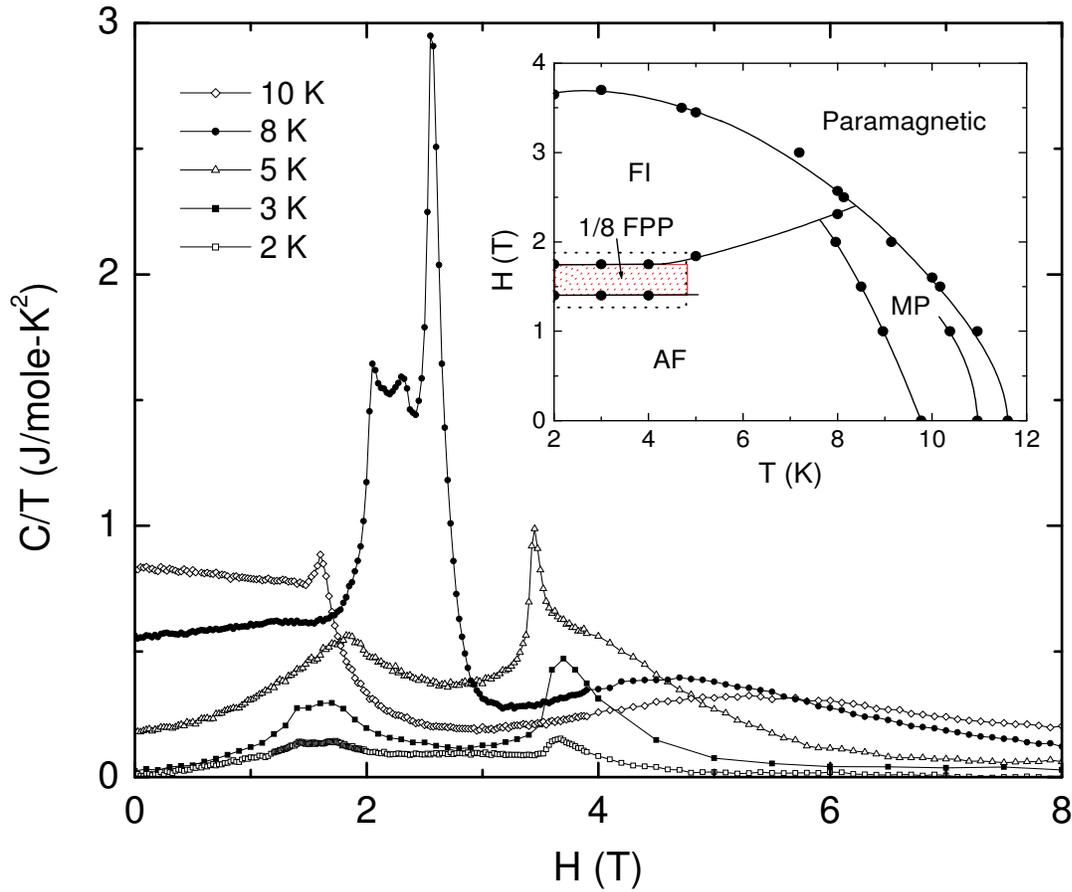

Fig. 1. Specific heat divided by temperature, *C*/*T*, as a function of magnetic field *H*, for various values of fixed *T*. Inset - Phase boundary as determined from these measurements as well as those of *C*/*T* vs. *T* at fixed *H* (available in supplemental information) showing the antiferromagnetic (AF), the ferrimagnetic (FI), the mixed phase (MP), and the 1/8 fractional plateau phase region (1/8 FPP).

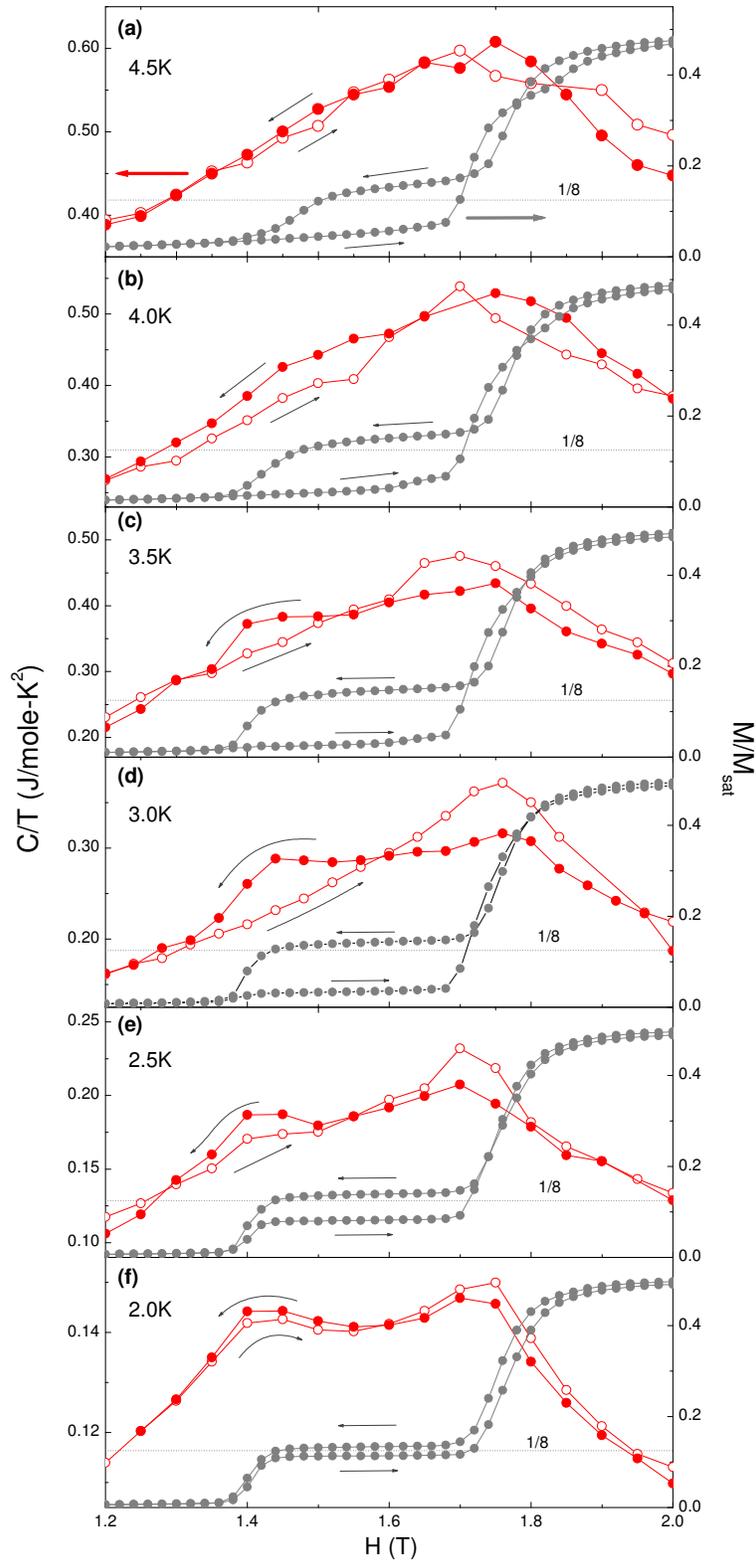

Fig. 2. Specific heat divided by temperature, *C/T*, and magnetization, *M*, as a function of magnetic field *H*, for various temperatures encompassing the FPP. The field sweep protocol is discussed in the text.

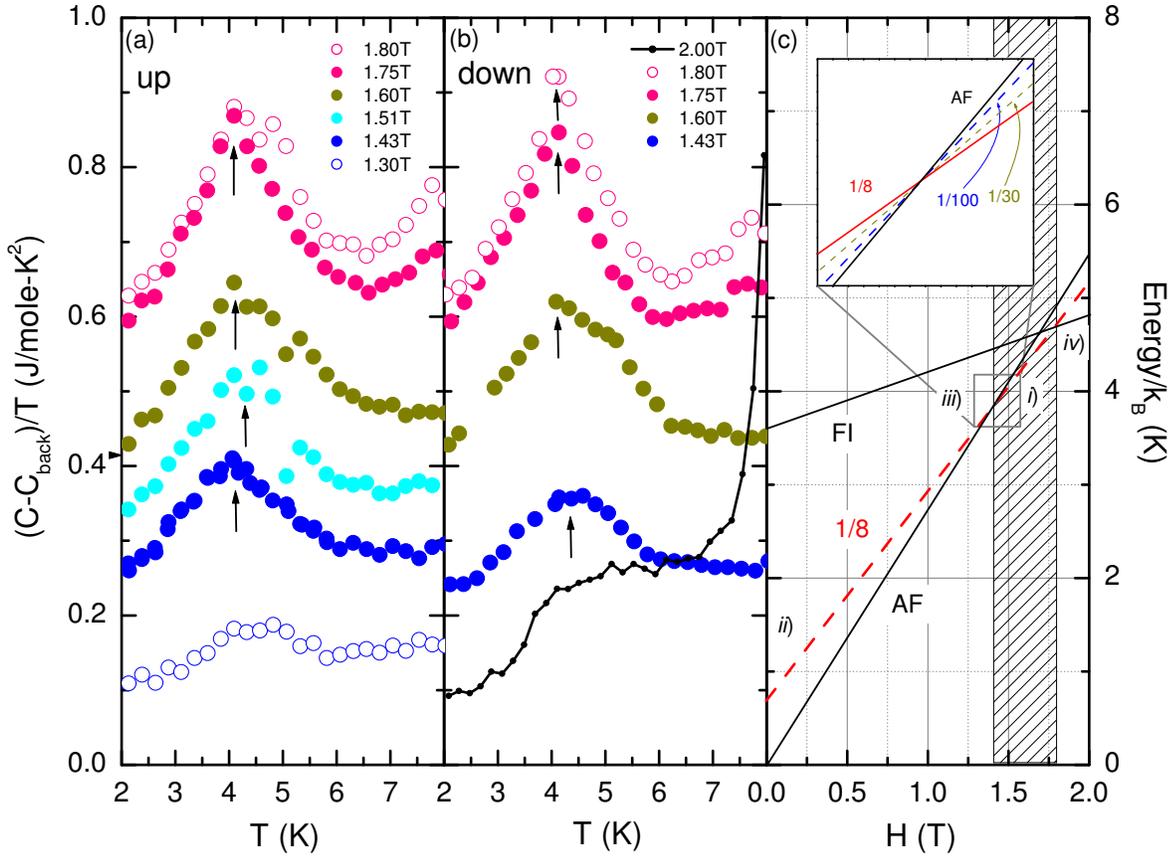

Fig. 3. (a) Specific heat divided by temperature, $C/T$, with a linear background term $C_{back} = 0.083(T-2)$, subtracted, versus temperature taken on increasing temperature, showing a peak that defines the high-temperature boundary of the FPP region. The data have also been offset by adding a constant equal to $(H-1.3)$. (b) Same as frame (a) but data taken on cooling. (c) Energy level diagram at zero temperature, as dictated by the positions of the phase boundaries via criteria *i*) - *iii*) as discussed in the text. Inset – Schematic of the expected energy levels of $M/M_s < 1/8$ plateau states.

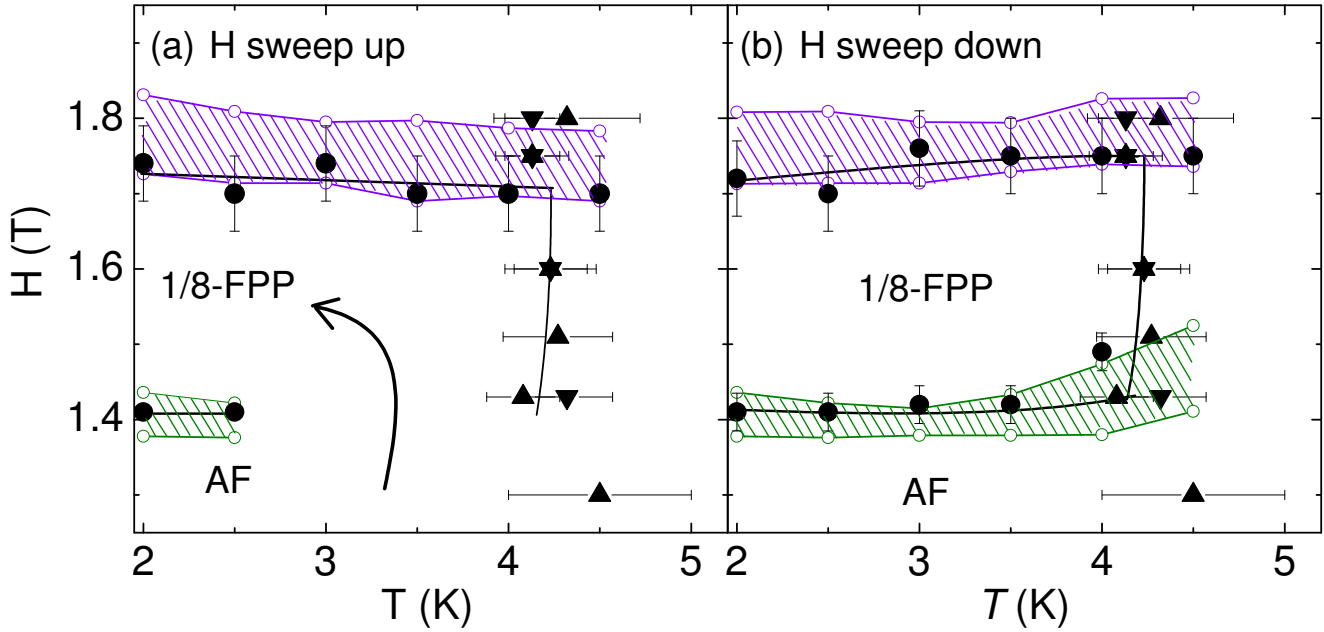

Fig. 4. The phase diagram around the 1/8 FPP as determined by specific heat measurements, denoted by solid black symbols. The circles are obtained from $C(H)$ in Fig. 2 and the up (down) triangles denote the peaks in $C(T)$ shown in Fig. 3a (b). The solid lines are guides to the eye. (a) Sweeping up in field - The low (high) field hatched regions are mixed 1/8-FPP/AF(FI) phases, as expected for first order transitions, defined by $M(H)$ measurements shown in Fig. 2. The curved arrow shows a possible route for entering the 1/8 FPP from the AF phase without crossing a phase boundary. (b) Sweeping down in field – The hatched regions are defined as in (a).